\documentclass[a4paper]{article}
\usepackage{algorithm2e}
\usepackage{amsmath}
\usepackage{amssymb}
\usepackage{graphicx}
\usepackage{float}
\usepackage{csquotes}
\usepackage{hyperref}
\usepackage{INTERSPEECH2022}
\usepackage{xcolor}
\title{A Passive Similarity based CNN Filter Pruning for Efficient Acoustic Scene Classification}
\name{Arshdeep Singh, Mark D. Plumbley}
\address{
 Centre for Vision, Speech and Signal Processing (CVSSP)\\
University of Surrey, UK
  }
\email{\{arshdeep.singh, m.plumbley\}@surrey.ac.uk}

\begin{document}
\RestyleAlgo{ruled}
\SetKwComment{Comment}{/* }{ */}
\maketitle
\begin{abstract}
We present a method to develop low-complexity convolutional neural networks (CNNs) for acoustic scene classification (ASC). The large size and high computational complexity of typical CNNs is a bottleneck for their  deployment on resource-constrained devices.
%ASC frameworks classify surrounding using sounds and are  deployed on resource-constrained devices. However, large size and computations in CNN is a bottleneck to deploy it on hardware.
We propose a passive filter pruning framework, where a few  convolutional filters from the CNNs are eliminated to yield compressed CNNs. Our hypothesis is that similar filters produce similar responses and give redundant information allowing such filters to be eliminated from the network. To identify similar filters, a cosine distance based greedy algorithm is proposed. A fine-tuning process is then performed to regain much of the performance lost due to filter elimination.
%However, the computations in the fine-tuning process are still there and directly depend on the number of training examples. 
To perform efficient fine-tuning, we analyze how the performance varies as the number of fine-tuning training examples changes.  An experimental evaluation of the proposed framework is performed on the publicly available DCASE 2021 Task 1A baseline network trained for ASC.  
The proposed method is simple, reduces computations per inference by  27\%, with 25\% fewer parameters, with less than 1\% drop in accuracy.

\end{abstract}

\section{Introduction}

Acoustic scene classification (ASC) frameworks  utilise sounds produced in the underlying environment to classify sounds into pre-defined scene classes \cite{barchiesi2015acoustic}. 
%These frameworks can be utilised for home monitoring, audio archive management and in  context aware services such as smart devices. 
Typically, ASC frameworks can be deployed on resource-constrained devices such as smart phones, particularly when millions of smart devices are connected in the network \cite{xu2008intelligent,temdee2018context,yang2021federated}. Therefore, reducing power consumption and memory storage of ASC frameworks is an important task.

Research in ASC actively started when the detection and classification of acoustic scenes and events (DCASE) community released the acoustic scenes dataset publicly and organised a challenge in 2013 \cite{Stowell2015}. Initially, DCASE challenges \cite{Mesaros2018_DCASE,Mesaros2018_TASLP,Mesaros2019_TASLP} focused on designing ASC frameworks to improve performance under different recording locations and using various recording devices. Recent DCASE challenges \cite{Heittola2020,martin2021low} for ASC have also focused  on designing low-complexity solutions to reduce computational complexity and memory storage.

Typically, learning-based methods have shown promising results compared to hand-crafted methods for ASC. Many learning-based methods use convolutional neural networks (CNNs) \cite{abesser2020review}, which give state-of-the-art performance in several related tasks such as audio classification \cite{kong2020panns} and image scene classification \cite{wang2020looking}. However, CNNs are resource hungry due to their large size and high computations \cite{denton2014exploiting}. Also, CNNs may have redundancy in their architecture  that lie in their redundant parameters that include weights or convolutional filters \cite{kahatapitiya2021exploiting}.

\begin{figure}[t]
    \centering
    \includegraphics[scale=0.4]{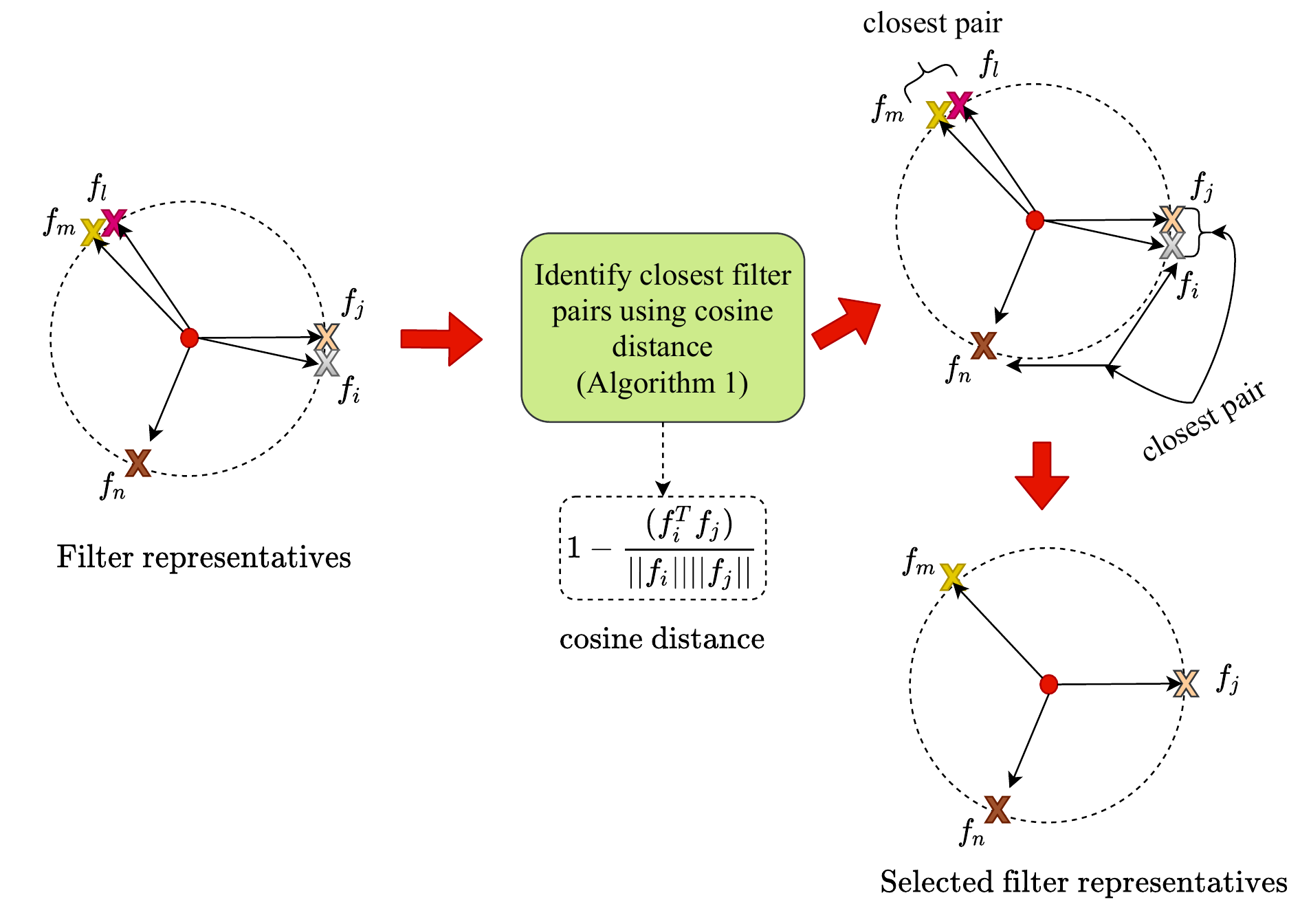}
    \caption{An illustration of a few important convolutional filters selected in a convolutional layer of CNN from a given set of filters using Algorithm 
    \ref{alg: identification of important filters}. Here, each filter is represented in a filter representative space.}
    \label{fig: Selection of imortant filters in a convolutional layer of CNN.}
\end{figure}

An illustration of similar convolutional filters where similarity is measured using cosine distance in filter representative space is shown in Figure  \ref{fig: Selection of imortant filters in a convolutional layer of CNN.}. Such similar filters mostly contribute to redundancy in CNNs \cite{singh2019deep} and hence, can be eliminated.     Eliminating redundancy in CNNs can reduce the number of parameters or memory storage and speed-up CNNs during their inference as well  \cite{kim2015compression}. In addition, the carbon footprint generated during training of CNNs reduce as training time per epoch decreases in compressed CNNs \cite{lacoste2019quantifying}. A few of the important convolutional filters selected using Algorithm \ref{alg: identification of important filters} proposed in Section \ref{sec: proposed methodology} are also shown in Figure \ref{fig: Selection of imortant filters in a convolutional layer of CNN.}.

The rest of this paper is organised as follows. Section \ref{sec: pruning methods} gives a brief overview on filter pruning methods to remove CNN parameters. The proposed framework to identify similar filters is described in Section \ref{sec: proposed methodology}. Next, Section  \ref{sec: performance eval} includes experimental setup and Section \ref{sec: results} presents results and analysis. Finally, the discussion and conclusion is presented in Section \ref{sec: discussion} and Section \ref{sec: conclusion} respectively.

%\begin{figure}[t]
 %   \includegraphics[scale=0.45]{interspeech_illustration.drawio.pdf}
 %   \vspace{-1.3cm}
 %   \caption{An illustration of Top-7 filters selected from (a) a set of 11 filters in a convolutional layer using  (b) the proposed method and (c) $l_1$-norm method \cite{li2016pruning}. The $l_1$-norm method eliminates filters with relatively low-norm, while our method ignores filters based on their similarity.}
  %  \label{fig: filter illiustration}
%\end{figure}

\section{Filter Pruning Methods}
\label{sec: pruning methods}

Filter pruning methods \cite{molchanov2016pruning, li2016pruning} have been widely employed to eliminate  convolutional filters  based on their importance and to yield a pruned network.  The importance  can be measured  either in an active manner that involves a dataset  \cite{singh2019deep,molchanov2016pruning,luo2017thinet,hu2016network,lin2019towards} or in a passive manner \cite{li2016pruning,zhuo2018scsp} that does not involve a dataset. Mostly, active filter pruning methods measure  entropy \cite{li2019using}, variance \cite{polyak2015channel} or similarity \cite{singh2019deep} in the outputs produced  by each filter across dataset to quantify their importance.  On the other hand, passive filter pruning methods \cite{li2016pruning, he2019filter}  rely only on convolution filters to quantify their importance  without involving any dataset, and hence are relatively simpler than active filter pruning methods.

Majority of the existing passive filter pruning methods are norm-based that uses e.g. $l_1$-norm \cite{li2016pruning}  of the filters to quantify their importance. A filter with relatively low norm is considered to be less important. After ranking filters based on their importance,  few filters are eliminated based on  a user-defined pruning ratio\footnote{The percentage of number of filters to be eliminated}. Eliminating filters reduce performance of the pruned network, which is regained by performing a fine-tuning step.

The efficacy of passive norm-based filter pruning methods in defining importance of each filter depend on two conditions: (a) the minimum norm of the filter should be close to zero, and (b) the difference between the minimum and maximum norm of the filters should be as large as possible. However, such conditions are not always satisfied \cite{park2020reprune}.

%The efficacy of such methods depends on two conditions: (a) the minimum norm of the filter should be close to zero, and (b) the difference between the minimum and maximum norm of the filters should be as large as possible. However, such conditions are not always satisfied \cite{park2020reprune}. 

%Moreover, norm-based methods do not consider the relationship among filters to quantify their importance and require a user-defined pruning ratio.

This paper proposes a passive filter pruning approach to compress any pre-trained convolutional neural networks designed for ASC. The proposed approach eliminates convolutional filters based on their similarity rather than considering filter norm, and does not require any user-defined pruning ratio.  Our hypothesis is that few of the  convolutional filters in CNNs are similar and such filters give mostly redundant information. Therefore,  one of the similar filters can be eliminated from the network without much loss of performance. In the proposed approach, we consider the relationship among convolutional filters by measuring their cosine-similarity to quantify their importance and proposes an Algorithm \ref{alg: identification of important filters} to select few important convolutional filters. 

%An illustration of the filters selected using the proposed method and $l_1$-norm method is shown in Figure \ref{fig: filter illiustration}. Eliminating similar filters from the network produces a pruned network with reduced parameters and computations per inference. 

We also analyse how many training examples are sufficient in the fine-tuning process to achieve similar performance to that obtained using all training examples.
%This helps in reducing computations during fine-tuning process are reduced due to pruned network and by reducing the training examples.
The major contributions of this paper are summarised as follows:

\begin{itemize}
    \item We propose a  passive filter pruning framework which removes similar filters from a pre-trained CNN without any requirement of user-defined pruning ratio.
    %by automatically selecting a pruning ratio.
   % \item A greedy algorithm is proposed to identify similar filters, and automatically select important filters without any requirement of user-defined pruning ratio.

    \item We show that a pruned network obtained using the proposed pruning  outperforms  an existing $l_1$-norm based pruning method.
    
    \item We show that a reduction of training examples by 25\%  during fine-tuning process gives a similar performance with reduced training time.
\end{itemize}

\section{Proposed Method}
\label{sec: proposed methodology}

Consider a convolutional layer in a  CNN which has $n$ filters denoted by $F_i$, $1 \le i \le n$. Each filter has the same size with width $w$, height $h$ and number of channels $c$. Given the $n$ filters from the convolutional layer, our aim is to identify similar filters and select few important filters.

First, each filter is transformed into a 2-D matrix of size ($wh \times c$), where each column is obtained by stacking the width and height parameters. Next, we find Rank-1 approximation of each filter by performing singular value decomposition (SVD) that gives best Rank-1 approximation with respect to the Frobenius norm.

A filter $F_i$ is approximated as $\hat{F}_i$ = $\sigma_1 u_1 v^T_1$, where $\sigma_1$ denotes the significant singular value, $u_1$ denotes first left  singular vector and $v_1$ denotes first right singular vector. 
Next, $\hat{F}_i$ is  normalised column-wise and  any column from the normalised $\hat{F}_i$ is chosen that acts as a \enquote{filter representative}, $f_i$. It is important to note that each column in the normalised $\hat{F}_i$ is same as $\hat{F}_i$ has unit rank.
%The above process is repeated for each filter in the given convolutional layer.
%An illustration to obtain filter representative  is shown in Figure \ref{fig: proposed method} (a). 
%\begin{figure}[h]
 %   \centering
  %  \includegraphics[scale=0.4]{AI4S_pruning-Page-1.pdf}
   % \vspace{-0.75cm}
%    \caption{An illustration to select important filters in a convolutional layer of a CNN.}
 %   \label{fig: proposed method}
%\end{figure}
After obtaining filter representatives corresponding to each filter, we compute the pairwise cosine distance between filter representatives. This gives a similarity matrix, $ \mathbf{W} \in \mathbb{R}^{n \times n}$. Now, given the similarity matrix, a greedy Algorithm \ref{alg: identification of important filters} is proposed to identify  important filters. 

To identify important filters,  we select the closest pair for each filter representative  using $\mathbf{W}$, and sorted each closest pair according to their distance in the order of low to high. Here, the cosine distance between the closest pair defines their importance.

Next, we define one of the filter representative as important or redundant from each closest pair starting from the least important closest filter pair. A filter representative (the first index) from the least important closest pair is chosen as an important and other as a redundant. This procedure is repeated for each closest pair in the order of their importance except for the closest filter pair containing already identified redundant filter representative. Finally, a set of  important filter representatives are obtained after ignoring all redundant filter representatives. An illustration of the important filters selected using Algorithm \ref{alg: identification of important filters} is shown in Figure \ref{fig: Selection of imortant filters in a convolutional layer of CNN.}.

%An illustration of  filters selected using the Algorithm \ref{alg: identification of important filters} is shown in Figure \ref{fig: proposed method}(b).

%Once important filters are obtained then the other filters are explicitly removed from the network to obtain a pruned network.

The redundant filters obtained using Algorithm \ref{alg: identification of important filters} are explicitly eliminated form the unpruned network to yield a pruned network. The pruned network thus obtained has reduced number of parameters, less computational complexity and loss in performance due to elimination of few filters. To regain lost performance, a fine-tuning step is performed which involves a training dataset to re-train the pruned network.
The proposed pruning algorithm can be found  at: \href{Our code is available at: https://gitlab.surrey.ac.uk/as0150/passive-pruning}{https://gitlab.surrey.ac.uk/as0150/passive-pruning}.

\begin{algorithm}[t]
\caption{Identification of important filters }
\KwData{Pair-wise cosine distance of $n$ filters \\ ($\mathbf{W}$  $ \in \mathbb{R}^{n \times n}$).}
\KwResult{Indices of important filters (Imp\_list)}
A= [\:], Imp\_list =[\:], Red\_list = [\:]\\
\For {$i \leq n$}{
  [$m$, $d$] = argmin\{ $\mathbf{W}$[ i, : - \{i\}] \} \textcolor{blue}{\Comment*[r]{Identify the closet filter $m$ to $i^{th}$ filter  with their distance $d$.}}
  A.append($(i,m)$, $d$); 
}
A\_sort = Sort(A)  \textcolor{blue}{\Comment*[r]{Sort $A$ based on the distance $d$.}}

\For {$i \leq len(\text{\textnormal{A}})$}{
   index\_imp = A\_sort[i][0]\textcolor{blue}{\Comment*[r]{important filter index}} 
   index\_red= A\_sort[i][1] \textcolor{blue}{ \Comment*[r]{redundant filter index}}
   \If{\text{\textnormal{index\_imp}} $\notin$ \text{\textnormal{Red\_list}}}{Imp\_list.append( index\_imp )\; Red\_list.append( index\_red )
   }
 } \label{alg: identification of important filters}
\end{algorithm}

\section{Experimental Setup}
\label{sec: performance eval}
We evaluate the proposed method on the  publicly available  DCASE 2021 Task1A baseline \cite{martin2021low} network to this we refer as the unpruned network that accepts log-mel energies of size $40 \times 500$  extracted from 10-second audio signals as input. The unpruned network has three convolutional layers (C1, C2, C3) with  16, 16 and 32 number of filters respectively. These are  followed by a dense layer of 100 units and a classification layer of 10 units.

The unpruned network is trained on TAU Urban Acoustic Scenes 2020 Mobile \cite{Heittola2020} development training dataset to classify 10 acoustic scenes recorded using multiple devices for  200 epochs using Adam optimizer with learning rate 0.001. The unpruned network gives 48.58\% accuracy on validation dataset, has 46246 parameters and requires 286M multiply-accumulate operations (MACs) per inference to produce an output corresponding to 10-second audio signal. For fine-tuning the pruned network, we opt similar conditions as used in training the unrpuned network except 30 fine-tuning epochs, which are reduced by  approximately 7x than that used in training the unpruned network.

%\noindent \textbf{Fine-tuning process:} After eliminating  the filters from the unpruned network, the pruned network is fine-tuned for 30 epochs with similar conditions as used to train the unpruned network.

%Training for more epochs consumes more training time and hence, more $\text{CO}_2$ emission \cite{lacoste2019quantifying}. This work intentionally reduces fine-tuning epochs (approximately seven times as that of training the unpruned network) with a motivation towards sustainable solutions to reduce $\text{CO}_2$ emission.

The effectiveness of the proposed pruning method is measured in terms of accuracy obtained with or without fine-tuning,  reduced MACs per inference and reduced  number of parameters in the pruned network. To report performance after fine-tuning, we fine-tune the pruned network independently for 5 times and reported average accuracy.

The proposed pruning method is compared to an $l_1$-norm based pruning method \cite{li2016pruning}. The $l_1$-norm based method ranks convolutional filters according to their norm and requires user-defined pruning ratio. Therefore, we choose the same pruning ratio for $l_1$-norm method as obtained using our proposed method. The experiments are performed on GeForce RTX 2070 with Max-Q Design computing device.

\section{Results \& Analysis}
\label{sec: results}

\begin{figure}[ht]
    \centering
    \includegraphics[scale=0.33]{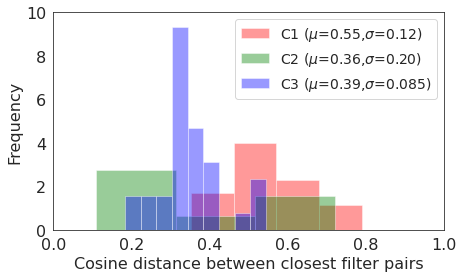}
    \caption{Histogram of distance between closest filter pairs across different convolutional layers. The mean $\mu$ and standard deviation $\sigma$  of cosine distance is also given.}
    \label{fig: histogram of smallest distance}
\end{figure}

The number of redundant filters obtained using the proposed pruning method in C1, C2 and C3 layers are 5, 5 and 10 respectively. The cosine distance of filters in C3  layer have relatively smaller deviation than that of other layers  as shown in Figure \ref{fig: histogram of smallest distance}. Due to this, the number of redundant filters in the C3 layer are more than that of other layers.

\begin{figure}[ht]
 
    \includegraphics[scale=0.42]{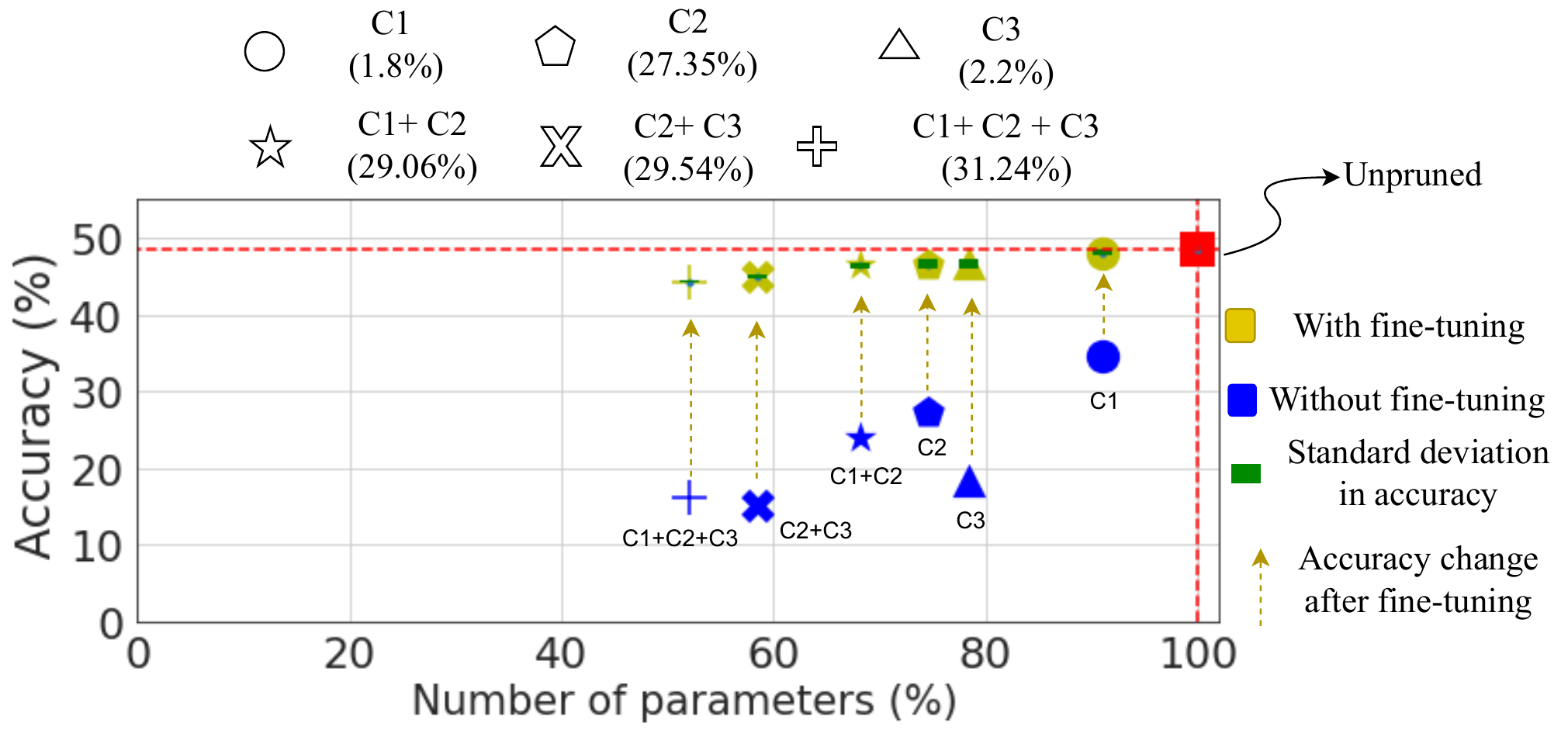}
    \vspace{-0.25cm}
    \caption{Various parameters obtained by pruning different layers in DCASE 2021 Task1A baseline network. Here, CX (\enquote{M}) means that the MACs have reduced by M\% after pruning X$^{th}$ convolutional layer. The accuracy is obtained with or without fine-tuning. The standard deviation in accuracy is obtained after fine-tuning the pruned network independently for 5 times using 100\% training dataset.} 
    %Here, the accuracy after fine-tuning of the pruned network is obtained by repeating same experiment for 5 times, and the average accuracy and standard deviation is reported. }
    \label{fig: various paramters after pruning different layers}
\end{figure}

Figure \ref{fig: various paramters after pruning different layers} shows the performance obtained after pruning different subset of layers. Without performing any fine-tuning, the accuracy in the pruned network degrades. Pruning C3 layer gives maximum reduction in accuracy in comparison to that of C1 and C2 layers. Also, pruning other layers in combination C3 layer shows relatively more degradation. We speculate that this is due to the significant alteration of decision boundaries learned by dense and classification layer as C3 is a bottleneck layer.

After performing fine-tuning, the accuracy of the pruned networks improve significantly as shown in Figure \ref{fig: various paramters after pruning different layers}. The accuracy obtained after fine-tuning C1 layer is similar to that of the unpruned network at 9\% reduction in parameters and 1.8\% reduction in MACs.  Even though C3 layer has more redundant filters yet C3 layer reduces only 2.2\% MACs with 22\% reduced parameters and   1\% drop in accuracy. On the other hand,  pruning C2 layer reduces 27\% MACs with 25\% reduction in parameters at less than 1\% drop in accuracy.  

%In the unpruned network, majority of the MACs occurs in C2 layer\footnote{It is important to note that the filter width and  filter height is same for all layers} as the size of the input to the C2 layer  is $16 \times (40 \times 500)$, which is larger than that of C1 layer ($1 \times (40 \times 500)$) and C2 layer ($16 \times (8 \times 100)$). Therefore, the C2 layer shows maximum reduction in MACs. 

In contrast to pruning individual layers, pruning two or more layers reduce more MACs and parameters at relatively larger drop in accuracy.  As an instance, pruning (C1+C2+C3) layers reduce approximately 31\% MACs and 48\% parameters at less than 4\% drop in accuracy.

%\noindent \textbf{Sensitivity analysis of various layers after pruning: Next, we analyse the sensitivity of each layers by measuring the accuracy obtained using the pruned network without performing any fine-tuning. Figure \ref{fig: pruned accuracy} shows accuracy for different layers. We observe that pruning C3 layer degrade the performance maximally.  This might be due to significant alteration of decision boundaries learned by dense and classification layer as C3 layer is a bottleneck layer.}

%\begin{figure}[h]
 %   \centering
 %   \includegraphics[scale=0.35]{Pruned_accuracy.png}
 
 %   \caption{Accuracy of the pruned network obtained after pruning different layers and without performing any fine-tuning.}
  %  \label{fig: pruned accuracy}
%\end{figure}
%As C3 layer is a bottleneck layer and removing filters from C3 layer affects the decision boundaries learned by dense and classification layer significantly.

\begin{figure}[ht]
    \centering
    \includegraphics[scale=0.4]{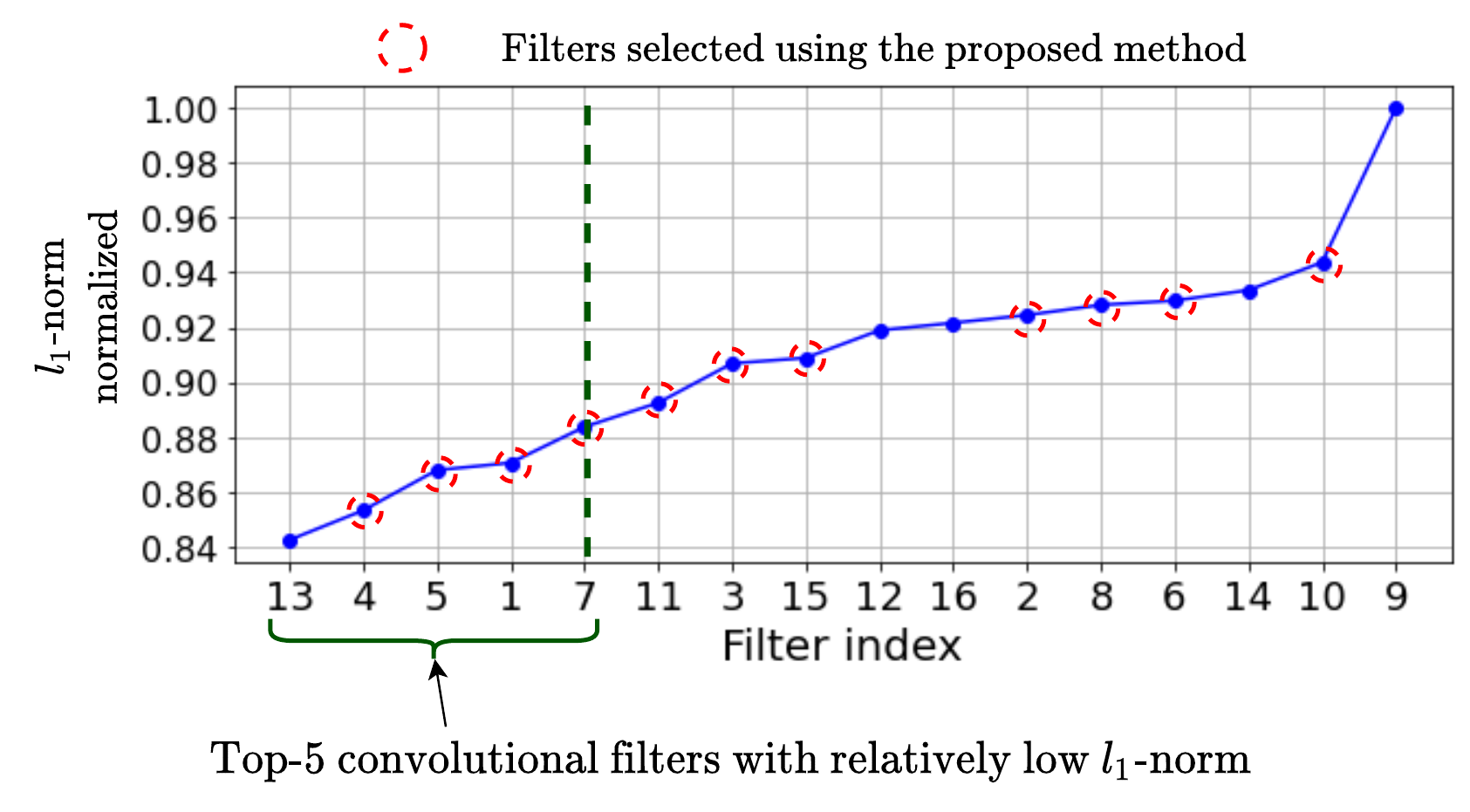}
        \vspace{-0.25cm}
    \caption{$l_1$-norm of the filters in C2 layer and top 11 filters selected using the proposed method. $l_1$-norm method ignores relatively low-norm filters indexed \{13, 4, 5, 1, 7\}.}
    \label{fig: Norm of filters}
\end{figure}

\noindent \textbf{Comparison with $l_1$-norm based pruning:}  The $l_1$-norm of the filters in C2 layer is shown in Figure \ref{fig: Norm of filters}. It can be observed that the top-5  filters with relatively low norm  may be still significant as the ratio of the lowest norm to highest norm is greater than 84\%. This verifies that the norm of the filter alone may not be a good metric to define filter importance. 
%The $l_1$-norm method always select relatively high norm filters as important and ignores relatively low norm filters as shown in Figure \ref{fig: Norm of filters}. 

On the other hand, our proposed method does not use  norm of the filters and rely on similarity among filters to select filters. It can be seen in Figure \ref{fig: Norm of filters} that our method selects filters with indices \{4,5,1,7\} which are ignored by $l_1$-norm method.

%Top-5 filters ignored by $l_1$-norm method are filters with indi$l_1$-norm method, the filters 

%The subset of filters selected in C2 layer are shown in Figure \ref{fig: Norm of filters}. The proposed method selects filters having relatively low-norm as well,  which are always ignored by $l_1$-norm method.

\begin{figure}[ht]
    \centering
    \includegraphics[scale=0.43]{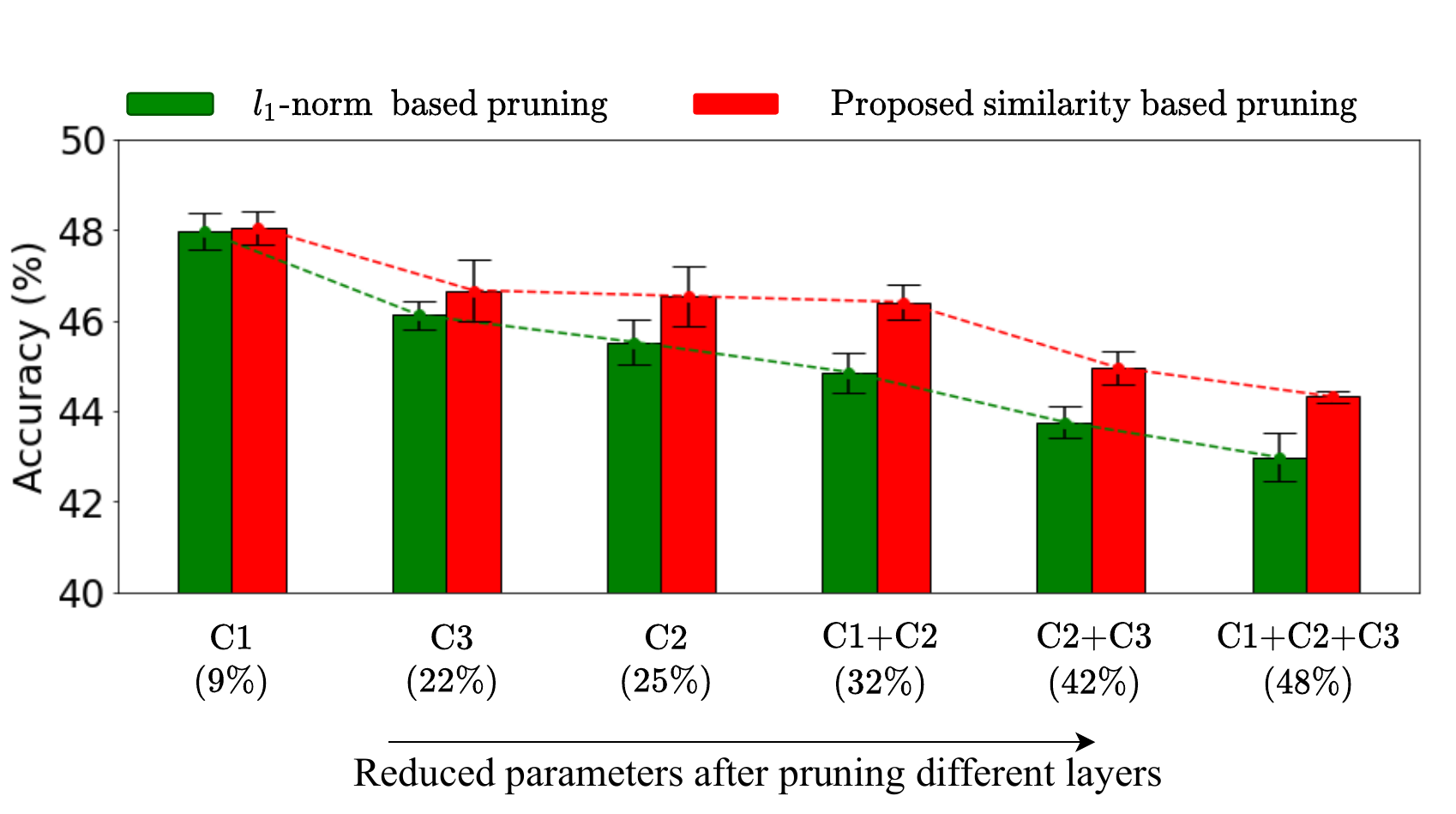}
    \vspace{-0.3cm}
    \caption{Accuracy obtained using proposed method and $l_1$-norm method after fine-tuning the pruned network using 100\% training dataset when different layers are pruned.}
    \label{fig: comparison of accuray}
\end{figure}

Choosing filters based on similarity improves performance of the pruned network as compared to norm-based method.  It can be observed from Figure \ref{fig: comparison of accuray} that the accuracy obtained using  the proposed similarity-based pruning  is better than that of  norm-based pruning across various layers. In particular, the proposed method improves accuracy by (1-1.5)\%, when the number of reduced parameters are more.

%\begin{table}[th]
%  \caption{Accuracy gain obtained using the proposed method over that of $l_1$-norm method after pruning various layers.} 
%  \vspace{-0.25cm}
  %Here, the experiment is repeated for 5 times to compute accuracy for each pruning method and the difference  between the average accuracy is reported. }
%  \label{tab: accuracy gain}
%  \centering
 % \begin{tabular}{lll}
 %   \toprule
 %  \textbf{Pruned layers} &  \textbf{Accuracy gain (\%)} &                                  \textbf{\# Parameters}\\
 %   \midrule
 %   No pruning & 0  & 46246 \\
%C1       & +0.07 & 42056 \\ 
%C2       & +1.01& 34461 \\ 
%C3       & +0.55 & 36256 \\ 
%C1+C2   & +1.55 & 31496 \\ 
%C2+C3    & (pending) & 27021 \\ 
%C1+C2+C3 & +1.48& 24056 \\ 
%    \bottomrule
%  \end{tabular}
  
%\end{table}

\begin{figure}[ht]
    \centering
    \includegraphics[scale=0.5]{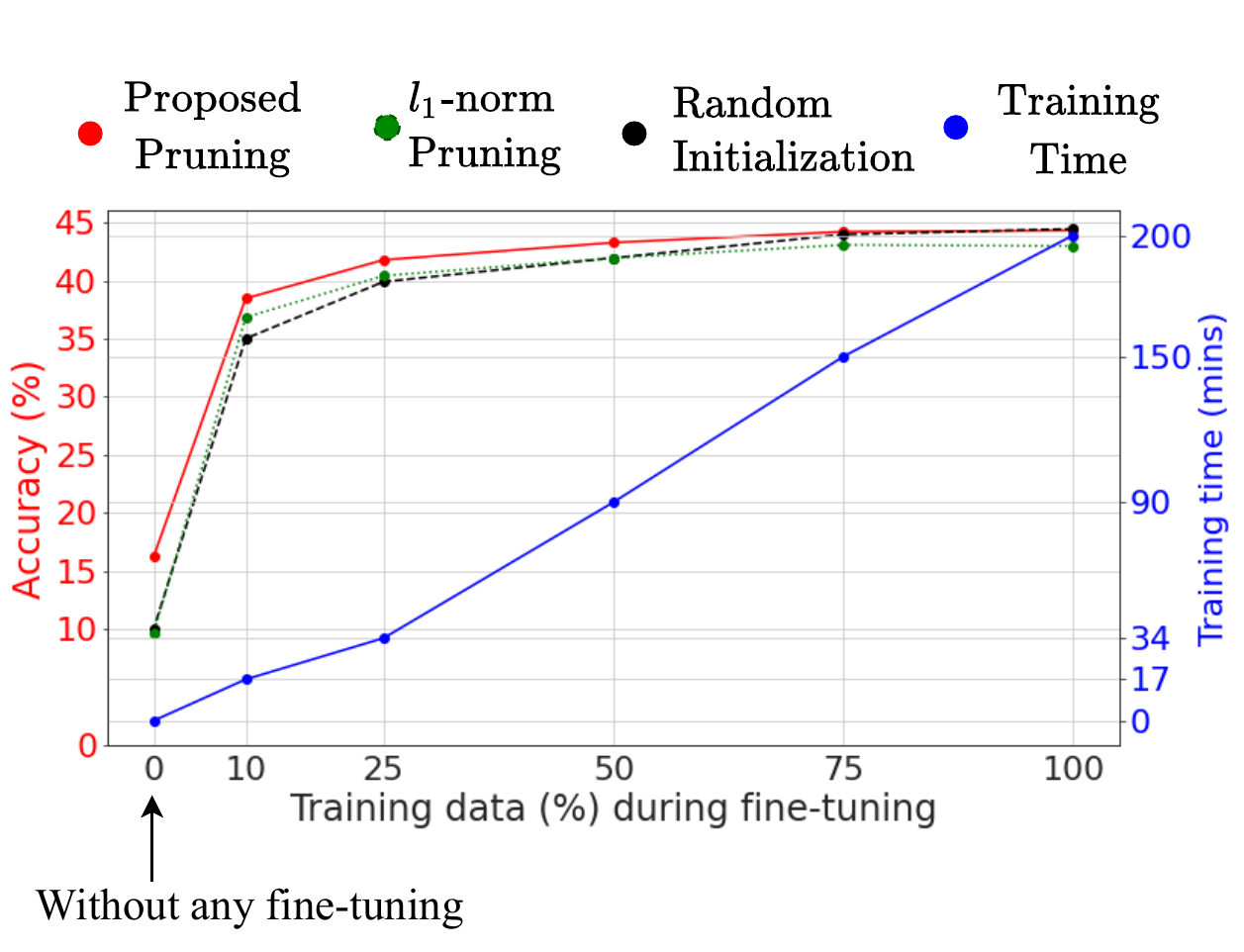}
    \vspace{-0.27cm}
    \caption{Accuracy and training time required to fine-tune by 30 epochs after pruning (C1 + C2 + C3) layers, when the number of training examples are varied from 10\% to 100\%.} 
    %For each training subset experiment, the fine-tuning process is repeated for 5 times and the average accuracy is reported.}
    \label{fig: accuracy and complexity}
\end{figure}

\noindent \textbf{Analysis of fine-tuning at different training examples:} Next, we analyse the computational complexity during fine-tuning and the accuracy obtained after fine-tuning  the pruned network by varying  training examples from 10\% to 100\% that is shown in Figure \ref{fig: accuracy and complexity}. The pruned network is obtained by pruning (C1+C2+C3) layers using the proposed method or $l_1$-norm method. Also, we fine-tune a pruned network where the parameters of the pruned network  are initialized randomly.

Without performing any fine-tuning, the accuracy of the pruned network obtained using the proposed pruning method is approximately 6 percentage points better than that of $l_1$-norm method or randomly initialized network. Also, the proposed pruning framework gives similar or better performance than $l_1$-norm method and randomly initialized network after performing fine-tuning. It is also interesting to note that the randomly initialized network performs similar to the pruned network obtained using the proposed pruning method when more training dataset ($\ge$ 75\%) is used in fine-tuning.

%The performance obtained using all training dataset for different pruned networks is more or less similar.

%This shows the efficacy of the proposed pruning method over that of  $l_1$-norm method and random initialization.  

The pruned network significantly improves  performance from 16\% to 40\% at 12x reduction in training time when fine-tuned with only 10\% training dataset in contrast to that of using all training dataset. Utilizing 75\% training dataset in fine-tuning the pruned network gives similar performance as that of all training dataset with 1.33x (25\%) reduction in training time.

%\noindent \textbf{Selecting filters by Chebyshev distance or by averaging filters:} We analyse the effectiveness of the proposed filter selection method by comparing the performance of the pruned network (obtained after pruning all layers), when the filters are selected using different distance metric such as Chebyshev. Also, we generate new filters by averaging the similar filters obtained from the proposed pruning method. 

\section{Discussion}
\label{sec: discussion}

The proposed pruning method produces a low-complexity pruned network that achieves similar performance as that of the unpruned network after performing fine-tuning for only 30 epochs. 

Pruning filters using the proposed method gives a set of important filters with improved performance in contrast to that of $l_1$-norm method. The proposed method is particularly advantageous when more parameters are eliminated. Moreover, the proposed method automatically selects number of important filters and does not require any user-defined pruning ratio.

\begin{table}[th]
  \caption{Accuracy of the pruned network obtained after pruning (C1+C2+C3) layers without performing any fine-tuning when different methods are used to select important filters.}    \vspace{-0.3cm}
  \label{tab: pruned network accuracy comparison}
  \centering
  \resizebox{0.45\textwidth}{!}{
  \begin{tabular}{ll}
    \toprule
   \textbf{Method to obtain Pruned network (C1+C2+C3)} &    \textbf{Accuracy}
                                          \\
    \midrule
    Proposed  cosine distance based  method & 16.23\% \\
    Chebyshev distance based method &  12.23\% \\
    Average filters & 10\% \\
    $l_1$-norm method & 9.84\% \\
    Randomly initialized pruned network & 10\%\\
    \bottomrule
  \end{tabular}}
\end{table}

%Typically, the pruned network fine-tuned for 30 epochs (7x less training time than that of the unpruned network) gives 1\% drop in accuracy as that of the unpruned network when appropriate layer (C2) is pruned at significant reduction in MACs and parameters. Pruning C1 layer gives similar accuracy to that of the unpruned network, however the reduced MACs are the smallest. The bottleneck layer C3 results in maximal accuracy drop after pruning. Pruning various subset of layers reduces MACs and parameters, however, the drop in accuracy increases.

We compare other methods such as Chebyshev distance and average filters  to compute important set of filters in the pruned network (C1+C2+C3) for analysing the effectiveness of the proposed pruning method. In Chebyshev distance method,  we use the Chebyshev distance which is defined as the maximum of  differences along any filter representative dimension to generate similarity matrix in Algorithm \ref{alg: identification of important filters}. In average filters method, we compute closest filter pairs based on cosine distance, and generate new filters by averaging the closest filter pairs.

The accuracy of the pruned network without performing any fine-tuning using the proposed method outperforms the other methods as given in Table \ref{tab: pruned network accuracy comparison}. The performance obtained after averaging the similar filters is more or less random. In contrast to Chebyshev distance, the cosine distance shows promising result.

%Apart from cosine distance based important filter selection, we choose filters using Chebyshev distance, where the distance between two filters is the maximum of their differences along any filter dimension, in Algorithm \ref{alg: identification of important filters} and also, we generate new filters by averaging two similar filters obtained from the cosine distance based Algorithm \ref{alg: identification of important filters}.  The comparison of performance of the pruned network with different filter selection criterion is shown in Table \ref{tab: pruned network accuracy comparison}. The filters selected using cosine distance based method performs better than that of other methods.

%A comparison of accuracy and computations (in terms of epochs) is given in Table xxx (ongoing).

\section{Conclusions and Future work}
\label{sec: conclusion}

We propose a cosine distance based passive filter pruning framework to compress a pre-trained convolutional neural network. The proposed framework identifies pruning ratio automatically.  We find that considering pair-wise relationships among filters yield a better set of important filters and improves performance compared to norm-based filters selection. We also observe that
fine-tuning the pruned network for few epochs and using few training examples improves the performance of the pruned network significantly.  Our future goal is  to design better distance metrics to measure pair-wise similarity among filters and to reduce the performance loss in the pruned network with minimal requirement of fine-tuning.

\section{Acknowledgements}

This work was supported by Engineering and Physical Sciences Research Council (EPSRC) Grant EP/T019751/1 \enquote{AI for Sound}.

%The ISCA Board would like to thank the organizing committees of the past INTERSPEECH conferences for their help and for kindly providing the template files. \\Note to authors: Authors should not use logos in the acknowledgement section; rather authors should acknowledge corporations by naming them only.

\newpage
\bibliographystyle{IEEEtran}

\bibliography{mybib}

% \begin{thebibliography}{9}
% \bibitem[1]{Davis80-COP}
%   S.\ B.\ Davis and P.\ Mermelstein,
%   ``Comparison of parametric representation for monosyllabic word recognition in continuously spoken sentences,''
%   \textit{IEEE Transactions on Acoustics, Speech and Signal Processing}, vol.~28, no.~4, pp.~357--366, 1980.
% \bibitem[2]{Rabiner89-ATO}
%   L.\ R.\ Rabiner,
%   ``A tutorial on hidden Markov models and selected applications in speech recognition,''
%   \textit{Proceedings of the IEEE}, vol.~77, no.~2, pp.~257-286, 1989.
% \bibitem[3]{Hastie09-TEO}
%   T.\ Hastie, R.\ Tibshirani, and J.\ Friedman,
%   \textit{The Elements of Statistical Learning -- Data Mining, Inference, and Prediction}.
%   New York: Springer, 2009.
% \bibitem[4]{YourName17-XXX}
%   F.\ Lastname1, F.\ Lastname2, and F.\ Lastname3,
%   ``Title of your INTERSPEECH 2022 publication,''
%   in \textit{Interspeech 2022 -- 23\textsuperscript{rd} Annual Conference of the International Speech Communication Association, September 18-22, Incheon, Korea, Proceedings, Proceedings}, 2022, pp.~100--104.
% \end{thebibliography}

\end{document}